\documentclass[11pt, twoside]{article}

\usepackage{asp2010}

\resetcounters

\begin{document}

	\title{Development of an Astrophysical Specific Language for Big Data Computation}
	\author{Nicolas KAMENNOFF$^1$, S\'ebastien FOUCAUD$^2$, and S\'ebastien REYBIER$^3$}
	\affil{$^1$Advanced Computer Science Epitech Laboratory (ACSEL) / Epitech, 14 rue Voltaire 94270 Le Kremlin-Bic\^etre, France}
	\affil{$^2$National Taiwan Normal University, Department of Earth Sciences, 88~Tingzhou~Road, Sec. 4, Wenshan district, Taipei 11677, Taiwan}
	\affil{$^3$Software and Mind Innovation (SoaMI), 5 all\'ee de la musique, 95210 Saint-Gratien, France}

	\begin{abstract}
		Astronomy is entering in a new era of Extreme Intensive Data Computation and we have identified three major issues the new generation of projects have to face: Resource optimization, Heterogeneous Software Ecosystem and Data Transfer.
		We propose in this article a middleware solution offering a very modular and maintainable system for data analysis.
		As computations must be designed and described by specialists in astronomy, we aim at defining a friendly specific programming language to enable coding of astrophysical problems abstracted from any computer science specific issues. 
		This way we expect substantial benefits in computing capabilities in data analysis. As a first development using our solution, we propose a cross-matching service for the Taiwan Extragalactic Astronomical Data Center.
	\end{abstract}

	\section{Context}
		Astronomy is already facing a massive data issue, while resources available to gather data overpass resources dedicated to their analysis.
		Future planned facilities such as the LSST and SKA will enhance even further this problem, producing public data at a pace never encountered before.
		Therefore such a challenge is also addressed to the Information Technology field, which is required to provide innovative solutions.
		As part of the setup of a new Data Center in Taiwan \citep{O14_adassxxii}, we aim at designing an open-source, distributed solution to enhance data analysis capabilities.

	\section{Selected Issues}
		\label{BLINK_ISSUES}
		Data analysis is presenting multiple contiguous issues, thus we are initially focusing on, what we think, the most beneficial to the problem of Big Data. We
		 divided them into three classes: Resource usage, heterogeneous software ecosystem and data transfer.\\
		
		\noindent{\bf $\bullet$ Resource usage}\\
			A vast majority of software were (and still are) developed considering that computing resources will increase indefinitely, without any focus on optimization.
			However, as we are facing a deluge of data, computation resources does not anymore follow the growth of our needs, as hardware is not evolving linearly. 
			Because of limitations due to electro-migration and sub-threshold conduction, the increase of processor speed has stopped in the past decade. Instead manufacturers multiplied processing units, leading to the current era of parallel computing, with multi-core and multi-processors CPUs, GPGPUs, APUs etc.
			This generalized modification of hardware requires a shift in our way of thinking software and most of the currently used software does not use efficiently all resources available.
			Creating efficient software, even using high quality framework, is a non trivial activity requiring appropriate (time-consuming) training, especially on low level layers computing. Software developers usually do not exploit efficiently the computing resources, because of unfortunate habits resulting from single core development, the misuse of CPU cache or usage of unfitted containers and data structures.\\

		\noindent{\bf $\bullet$ Heterogeneous software ecosystem}\\
			A diverse range of data analysis software are already available to the community for different mainstream tasks, the main part of them developed as standalone services. Researchers usually write scripts calling a chain of software, saving them the hurdle of developing of parts of the analysis system from scratch.
			However, using different software with unsuited communication system (using files or wordy protocols) reduces drastically performances.
			In fact, writing and reading files is  an inconsiderate bottleneck, as access to hard drives reduce drastically performances of the analysis tool chain.\\

		\noindent{\bf $\bullet$ Data transfer}\\
			The Virtual Observatory (VO) has been developed to exchange easily public data stored in different Data Centers. In this configuration, accessing to data or part of the data usually requires to use the Data Access Layer and store the outputs locally.
			Therefore each analysis will require a partial download of the data, resulting in a overhead during the computation, which can severely affect performances.
			Depending on the local Internet download speed, the VO upload capabilities and usage, manipulating vast amount of data represents a major issue, especially for small research laboratories, which may require important (and costly) amount of storage.
One obvious solution to this problem is to move towards the data location, therefore develop data analyses services hosted by the Data Center themselves to which users can connect remotely.

	\section{State of the Art}
		These problems are known by the astronomical community and several solutions have already been proposed. We here focus on two recent attempts: SAMP and FASE.\\
		
		\noindent{\bf $\bullet$ Simple Application Messaging Protocol (SAMP)}\\
			SAMP \citep{samp} is a solution brought by the Application Group of Interest from the IVOA consortium. 
			SAMP defines a messaging protocol enabling easy interoperability between the different analysis software and reducing overhead due to file transfers.
			The IVOA currently recognized that an attempt to build a monolithic tool is not a rational solution. SAMP then defines how applications should be able to collaborate and share their data, bringing some partial solution to the data transfer and heterogeneous environment issues.
			However, this system does not handle data localization nor interfere with optimization of resource usage from each software.
			Indeed, as IVOA advocate for an interaction of various software, development specific issues are not addressed. \\
			
		\noindent{\bf $\bullet$ Future Astronomical Software Environment (FASE)}\\
			FASE \citep{fase1} aims to enable analysis software in a shared environment system, using VO protocols for the software to interoperate (SAMP).  The project has been defined within the OPTICON  Network (Optical-Infrared Co-ordination Network for Astronomy) funded by the european FP7 program.  FASE is still actually a prototype for proof-of-concept and its first goal has been to define packaging requirements for the analysis software to be distributed within the same environment. In a second phase, \cite{fase2} emphasize the accessibility of FASE to legacy software as well as its capability in enabling user scripting routines.

	\section{Toward a Domain Specific Language}
		Working toward an unified ecosystem for astrophysical data analysis will be unavoidable very soon.
		The current solutions (SAMP and FASE) only address part of the main issues emphasized in section~\ref{BLINK_ISSUES}. Obviously a monolithic software approach does not make sense, but we advocate here that a modular distributed middleware is a valuable solution.
		Considering the limited resources available to develop sustainable features and softwares in astronomy, sharing common parts of algorithms and data structures used by the different software is essential. Furthermore Astronomers do not always have sufficient training to deal with low level layers programming, which should be developed by IT specialists. We therefore propose to explore a Domain Specific Language for astronomical data analysis.\\

		\noindent{\bf $\bullet$ Using every resources available}\\
			As improvement of the available hardware is reaching a physical and financial limit, it is important to maximize the use of all available resources.
			Recent tools and frameworks allows easy programming on multi-core CPUs, GPGPUs and APUs, but usually at the cost of efficiency. The solution we are exploring is to adapt the pool of algorithms used accordingly to the type of hardware available.\\

		\noindent{\bf $\bullet$ Strong Scheduling to create efficient computation pipelines}\\
			As available resources can be distant or already used by another computation, one key element is to define a robust scheduling system.
			Such a system will be able to identify available resources dynamically and select them for a specific request based on various characteristics: Distance from the data, Computation and Memory capabilities, Availability, Reliability, etc.
			As we intend to offer a system easily deployable, we have to design a strong scheduling service able to learn and monitor its own resources, using graph learning, graph mining and decision machine learning. Such a system would also be able to run simulation test.
			Also, we plan to use data mining and input categorization to enable the scheduling service in efficiently choosing the way the request is computed according to the request itself or the data required.\\

		\noindent{\bf $\bullet$ Dynamic Software Product Line for strong modularity and good performances}\\
			Dynamic Software Product Line \citep[DSPL -][]{DSPL} is a rising paradigm in Computer Science which focus on adaptation capabilities of the software during runtime instead of forging it during compilation.
			This could allow to adapt a running software to fit the new computation needs and still reach near optimal performances.
			This way, we can create and adapt computation pipelines on-the-fly which will fulfill users requests as fast as possible.\\
		
		\noindent{\bf $\bullet$ In Memory Processing}\\
			Input and Output (I/O) latency, such as getting, storing and updating information on a hard drive, is a strong disabling bottleneck. By building a uniform and strongly scheduled pipeline, we can ensure that the requested computations are mainly performed in-memory, as Random Access Memory (RAM) is far more efficient than I/O.\\
						
		\noindent{\bf $\bullet$ Next steps and expected roadmap}\\
			We are currently developing on the first tool using our framework for CPU, GPGPU and APU architectures. This HTM Quadtree algorithm is part of a cross-matching service \citep[BLINK -][]{blink} that we will provide through the Taiwan Extragalactic Astronomical Data Center \citep[TWEA-DC -][]{O14_adassxxii} from spring 2013.
			In the mean time we are designing a standalone service to set on computation resources and the basic features of the scheduling service.
			We are also leading a Space Partitioning Survey on GPGPU with American Micro Device (AMD) which should be released next summer.
			As it aims to design a new simplified programming language for the astronomical research area, the project requires a close collaboration between astronomers and IT specialists. Discussing the DSL specification, format and important first features is essential and we are planning to release by the end of the year a dedicated website and community tools. We invite interested parties to check as we will publish informations on the standard mailing lists soon.

\bibliography{O18}

\end{document}